\documentstyle[prl,eqsecnum,aps,floats,epsfig,amssymb]{revtex} 

\def\fsec   {fs}                                

\def\dztau  {407.9}                             
\def\dzetau {6.0}                               
\def\dzstau {4.3}                               
\def\dznev  {10210}                             
\def\dzenev {125}                               

\def\lctau  {198.1}                             
\def\lcetau {7.0}                               
\def\lcstau {5.6}                               
\def\lcnev  {1630}                              
\def\lcenev {45}                                

\def\dzkitau  {416}                             
\def\dzkietau {12}                              
\def\dzkinev  {2470}                            
\def\dzkienev {57}                              

\def\dzbkitau  {416}                            
\def\dzbkietau {10}                             
\def\dzbkinev  {3420}                           
\def\dzbkienev {65}                             

\def\dzkiiitau  {399}                           
\def\dzkiiietau {16}                            
\def\dzkiiinev  {1950}                          
\def\dzkiiienev {63}                            

\def\dzbkiiitau  {400}                          
\def\dzbkiiietau {14}                           
\def\dzbkiiinev  {2360}                         
\def\dzbkiiienev {66}                           

\def\dzavrtau  {410.3}                          
\def\dzavretau {6.3}                            
\def\dzavrchi  {0.49}                           

\def\dpkiitau  {1070}                           
\def\dpkiietau {36}                             

\def\kzii {{\mbox{$K_s^0 \rightarrow \pi^+ \pi^- \ $}}} \def\kiii
{{\mbox{$K^-\pi^+\pi^-\pi^+ \ $}}} \def\dzki {{\mbox{$D^{0} \rightarrow
      K^-\pi^+ \ $}}} \def\dzbki {{\mbox{$\overline{D}^{0} \rightarrow
      K^+\pi^- \ $}}} \def\dzkiii {{\mbox{$D^{0} \rightarrow
      K^-\pi^+\pi^-\pi^+ \ $}}} \def\dzbkiii {{\mbox{$\overline{D}^{0}
      \rightarrow K^+\pi^-\pi^+\pi^- \ $}}} \def\lcpki
{{\mbox{$\Lambda_{c}^{+} \rightarrow pK^-\pi^+ \ $}}}

\begin{document}

\title{Precision measurements of the $\Lambda_{c}^{+}$ and $D^{0}$ lifetimes}

\author{
A.~Kushnirenko$^{3}$,
G.~Alkhazov$^{11}$,
A.G.~Atamantchouk$^{11}$,
M.Y.~Balatz$^{8}$$^{,\ast}$,
N.F.~Bondar$^{11}$,
P.S.~Cooper$^{5}$,
L.J.~Dauwe$^{17}$,
G.V.~Davidenko$^{8}$,
U.~Dersch$^{9}$$^{,\dag}$,
A.G.~Dolgolenko$^{8}$
% ...
\\                                                                            
\vskip 0.50cm                                                                 
\centerline{(SELEX Collaboration)}   
\vskip 0.50cm                                                                 
}

\date{\today}
\address{
$^1$Ball State University, Muncie, IN 47306, U.S.A.\\
$^2$Bogazici University, Bebek 80815 Istanbul, Turkey\\
$^3$Carnegie-Mellon University, Pittsburgh, PA 15213, U.S.A.\\
$^4$Centro Brasiliero de Pesquisas F\'{\i}sicas, Rio de Janeiro, Brazil\\
$^5$Fermilab, Batavia, IL 60510, U.S.A.\\
$^6$Institute for High Energy Physics, Protvino, Russia\\
$^7$Institute of High Energy Physics, Beijing, P.R. China\\
$^8$Institute of Theoretical and Experimental Physics, Moscow, Russia\\
$^9$Max-Planck-Institut f\"ur Kernphysik, 69117 Heidelberg, Germany\\
$^{10}$Moscow State University, Moscow, Russia\\
$^{11}$Petersburg Nuclear Physics Institute, St. Petersburg, Russia\\
$^{12}$Tel Aviv University, 69978 Ramat Aviv, Israel\\
$^{13}$Universidad Aut\'onoma de San Luis Potos\'{\i}, San Luis Potos\'{\i}, Mexico\\
$^{14}$Universidade Federal da Para\'{\i}ba, Para\'{\i}ba, Brazil\\
$^{15}$University of Bristol, Bristol BS8~1TL, United Kingdom\\
$^{16}$University of Iowa, Iowa City, IA 52242, U.S.A.\\
$^{17}$University of Michigan-Flint, Flint, MI 48502, U.S.A.\\
$^{18}$University of Rome ``La Sapienza'' and INFN, Rome, Italy\\
$^{19}$University of S\~ao Paulo, S\~ao Paulo, Brazil\\
$^{20}$University of Trieste and INFN, Trieste, Italy\\
}
\maketitle

\begin{abstract}
  We report new precision measurements of the lifetimes of the
  $\Lambda_{c}^{+}$ and $D^{0}$ from SELEX, the charm hadro-production
  experiment at Fermilab.  Based upon \lcnev~$\Lambda_{c}^{+}$ and
  \dznev~$D^{0}$ decays we observe lifetimes of $\tau[\Lambda_{c}^{+}] =
  \lctau \pm \lcetau \pm \lcstau$~{\fsec} and $\tau[D^{0}] = \dztau \pm
  \dzetau \pm \dzstau$~{\fsec}.
\end{abstract}

\twocolumn

Lifetime measurements of the charm baryons help to determine the contributions
of non-spectator weak interaction effects like W-annihilation and W-exchange
processes without the helicity suppression that limits their role in charm
meson decays.  From the point of view of Heavy Quark Effective Theory and
Perturbative QCD, the charm baryon lifetimes can be expressed in terms of a
set of matrix elements that contain the corrections to the fundamental
expansion of the decay amplitude in terms of
1/$m_c$~\cite{Korner,HQEH,Bellini:1997ra}.  The $\Lambda_{c}^{+}$ lifetime is
the best-measured of the four stable charm baryons~\cite{PDG}.  We present a
new measurement from hadro-production data taken by the
SELEX(E781)~\cite{SELEX} experiment at Fermilab.  Using the same data sample,
cuts and techniques, we have also measured the lifetime of the $D^{0}$ with a
precision comparable to the best present measurements~\cite{PDG}.  This new
$D^{0}$ measurement verifies our lifetime analysis procedure in a sample with
higher statistical precision and larger corrections than the
$\Lambda_{c}^{+}$.  Details may be found in ref~\cite{Thesis}.

\begin{figure}[ht]
\centerline{\psfig{figure=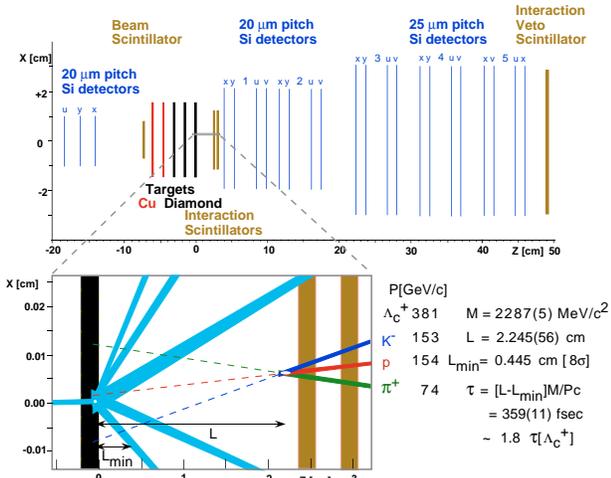,width=80mm}}
\caption{The charm targets and vertex detector.   A clear example of a $\Lambda_{c}^{+}$ event with track error corridors and vertex error ellipses is shown in the expanded region.}
\label{fig:apparatus}
\end{figure}

The SELEX experiment uses the Fermilab charged hyperon beam at 600 GeV to
produce charm particles in a set of thin foil targets of Cu or diamond.  The
three-stage magnetic spectrometer is shown elsewhere~\cite{Thesis,SELEX}.  The
most important features for the charm lifetime studies are the high-precision
vertex detector that provides an average proper time resolution of 20~{\fsec}
for the charm decays, a 10 m long Ring-Imaging Cerenkov (RICH) detector that
separates $\pi$ from K up to 165 GeV/c~\cite{RICH}, and a high-resolution
tracking system that has momentum resolution of ${\sigma}_{P}/P<1\%$ for a
\mbox{200\,GeV/c} reconstructed $\Lambda_{c}^{+}$.  Figure \ref{fig:apparatus}
shows the vertex region in detail with an overlay of reconstructed tracks,
error corridors and measured parameters for a clear $\Lambda_{c}^{+}$event.

The experiment selected charm candidate events using an online
secondary vertex algorithm.  A scintillator trigger demanded an
inelastic collision with at least four charged tracks in the
interaction scintillators and at least two hits in the positive
particle hodoscope after the second analyzing magnet.  Event selection
in the online filter required full track reconstruction for measured
fast tracks ($p\,{\scriptstyle\gtrsim}\,15\,\mathrm{GeV}/c$).  These
tracks were extrapolated back into the vertex silicon planes and
linked to silicon hits.  The beam track was measured in upstream
silicon detectors.  A full three-dimensional vertex fit was then
performed.  An event was written to tape if all the fast tracks in the
event were \emph{inconsistent} with having come from a single primary
vertex.  This filter passed 1/8 of all interaction triggers and had
about $50\%$ efficiency for otherwise accepted charm decays.  The
experiment recorded data from $15.2 \times 10^{9}$ inelastic
interactions and wrote $1 \times 10^{9}$ events to tape using both
positive and negative beams. $65\%$ of events were $\Sigma^{-}$
induced with the balance split roughly equally between $\pi^{-}$ and
protons.

The analysis selected charm events with a topological identification
procedure.  Only charged tracks with reconstructed momenta were used.
Tracks which traversed the RICH
($p\,{\scriptstyle\gtrsim}\,22\,\mathrm{GeV}/c$) were identified as
protons or kaons if those hypotheses were more likely than the pion
hypothesis.  All other tracks were assumed to be pions.  The primary
vertex was refit using all found tracks.  An event was rejected if all
the tracks were consistent with only a primary vertex.  For those
which were inconsistent, secondary vertices were formed geometrically
and then tested against a set of charge, RICH-identified and mass
conditions to identify candidates for the different charm states.

\begin{figure}[ht]
\centerline{\psfig{figure=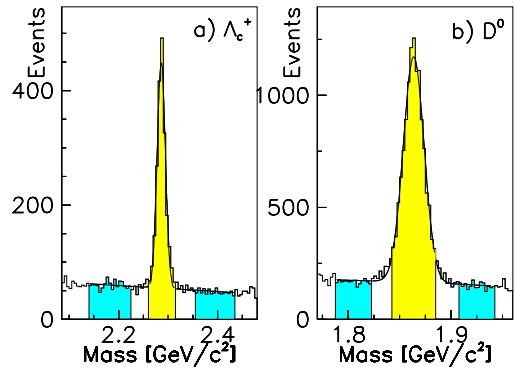,width=80mm}}
\caption{The mass distribution for the 
  a) $\Lambda_{c}^{+}$ sample in 5~MeV/$c^2$ bins and  
  b) $D^{0}$ sample in 2.5~MeV/$c^2$ bins. 
The signal and sideband regions are shaded.}
\label{fig:mass}
\end{figure}

The charm decay modes used were $\lcpki$ and $\dzki, K^-\pi^+\pi^-\pi^+$ +
charge conjugate.  No $\overline{\Lambda_{c}}^{-}$ candidates were considered
because of the strong production asymmetry in this data sample.  The charm
event selection criteria required: i) acceptable fits for all tracks and
vertices, ii) all track momenta exceed 8 GeV/c, iii) proton and kaon tracks to
be RICH-identified, iv) the secondary vertex to reconstruct upstream of the
interaction counters and at least 0.5 mm from any target or other material, v)
the significance of the primary-secondary vertex separation, L, be at least
8$\sigma$, where $\sigma$ is the error on L, vi) $\sigma$ to be less than 1.7
mm, vii) two charm decay tracks with miss distances to the primary vertex
greater than 20 $\mu$m in space, viii) and the charm momentum be parallel to
the vector from primary to secondary vertex within errors.  The mass peaks for
the candidate events selected are shown in Fig.~\ref{fig:mass}.

Because the proper time resolution is short compared to the expected 
$\Lambda_{c}^{+}$ lifetime of $\sim$ 200~{\fsec}, we use a binned 
lifetime analysis.  We bin in reduced proper lifetime; $t_{R}=[L-L_{min}]M/Pc$,
where M is the known charm state mass~\cite{PDG}, P its reconstructed 
momentum, L the measured vertex separation and $L_{min}$ the minimum L for 
each event to pass all the imposed selection cuts. $L_{min}$ varies event by
event.  This quantity $t_R$ should have an exponential distribution with the lifetime of the decaying state for 
acceptance-corrected signal events.

To correct the raw proper time distributions, one must understand the
apparatus acceptance as a function of the proper time.  Apparatus acceptance
for a charm decay at a given proper time depends on event variables: momenta,
decay configuration, position along the axis of the apparatus, and track
multiplicity.  A suitable simulation program would not only produce
correctly the kinematics of charm pair production but also have a correct
reproduction of the underlying event.  Because neither the true distributions
of track characteristics in the underlying event nor the true production
properties of charm hadrons in our data (momentum, track multiplicities
\ldots) are known, we decided to evaluate the proper time acceptance for the
sample of events {\it that we actually observe}.  In the SELEX apparatus,
proper time acceptance depends only on the vertex region detectors.
Downstream detectors could not resolve shifts of the decay vertex.  Each event
was re-analyzed by moving the charm decay point to different distances $L$
from the primary vertex. The event topology, momenta and other properties of
the event were kept fixed.  The analysis code was then run to decide if a
charm decay at this particular distance $L$ would be accepted or rejected.  In
such a way each individual event efficiency as a function of reduced proper
time $t_{R}$ was formed.  The overall efficiency of the observed sample is
just the weighted average of the individual event efficiencies.  This
technique preserves the production and acceptance properties and correlations
in the data including the underlying event without requiring a complete
simulation of charm production.

\begin{figure}[!ht]
\centerline{\psfig{figure=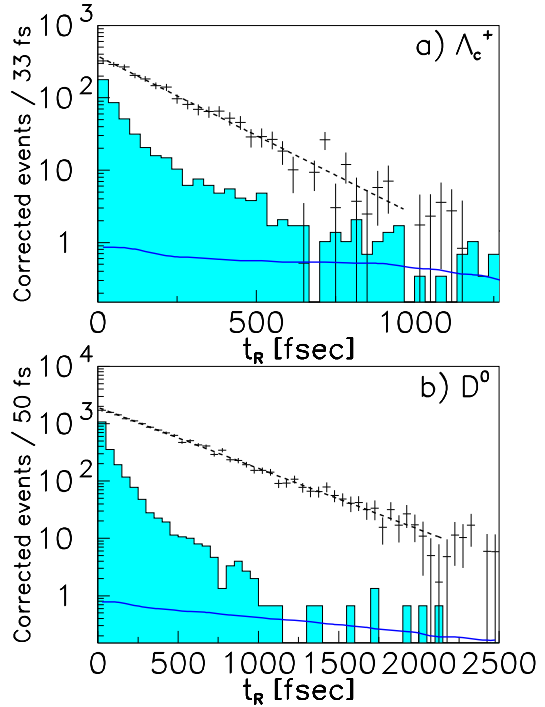,width=80mm}}
\caption{The acceptance-corrected reduced proper lifetime 
  distributions for the background subtracted signal (points) and 
  sideband (shaded) regions for 
  a) $\Lambda_{c}^{+}$ in 33~{\fsec} bins and
  b) $D^{0}$ in 50~{\fsec} bins. 
  The dashed line is the lifetime fit.
  The background is normalized to the width of the signal region shown in
  Fig.~\ref{fig:mass}.  The solid line is the acceptance as a function of
  $t_{R}$.}
\label{fig:lifetime}
\end{figure}

We make $t_{R}$ distributions for the signal and sideband regions, shown
in Fig.~\ref{fig:mass}.  A simultaneous maximum likelihood fit to both
the signal and sideband distributions is made.  The sideband distribution is
represented with a background function (the sum of two exponentials times
acceptance).  The signal distribution is represented with the same background
function plus an exponential times acceptance for the lifetime.  The
acceptance, acceptance-corrected distributions and fits are shown in
Fig.~\ref{fig:lifetime}.

\begin{table}[!ht]
\centering
\begin{tabular}{|l||c|c|}
Decay Mode              & $\tau$ ({\fsec})    & Events~~         \\ 
\hline \hline
\dzki,\kiii +cc         & $\dztau \pm \dzetau$  & $\dznev \pm \dzenev$  \\ 
\lcpki                  & $\lctau \pm \lcetau$  & $\lcnev \pm \lcenev$  \\ 
\hline \hline
\dzki                   &   $\dzkitau \pm   \dzkietau$ &   $\dzkinev \pm   \dzkienev$   \\ 
\dzbki                  &  $\dzbkitau \pm  \dzbkietau$ &  $\dzbkinev \pm  \dzbkienev$   \\ 
\dzkiii                 & $\dzkiiitau \pm \dzkiiietau$ & $\dzkiiinev \pm \dzkiiienev$   \\ 
\dzbkiii                &$\dzbkiiitau \pm\dzbkiiietau$ &$\dzbkiiinev \pm\dzbkiiienev$   \\ 
\hline
average                 & $\dzavrtau \pm \dzavretau$  &                  \\ 
$\chi^{2}/dof$          & \dzavrchi                   &                  \\   
\hline \hline
Target                  & $\Lambda_{c}^{+}$\hspace{1.0ex}$\tau$({\fsec}) 
                        & $D^{0}$\hspace{1.0ex}$\tau$ ({\fsec})          \\
\hline \hline

%

         1  Copper      & $198 \pm 20$     & $394  \pm 13$     \\
         2  Copper      & $198 \pm 22$     & $422  \pm 14$     \\
         3  Diamond     & $229 \pm 25$     & $413  \pm 15$     \\
         4  Diamond     & $178 \pm 14$     & $412  \pm 14$     \\  
         5  Diamond     & $202 \pm 16$     & $413  \pm 16$     \\  
\hline                                                         
average                 & $195.2 \pm 8.2$  & $410.1 \pm 6.4$   \\ 
$\chi^{2}/dof$          & 0.88             & 0.55              \\ 
\end{tabular}
\caption{Complete and sub-sample lifetimes with statistical errors.}
\label{tab:lifetimes}
\end{table}

As a consistency check we have repeated the analysis for
each decay mode and for events from each target separately.  The acceptance 
function changes significantly between these cases.  The lifetimes from these 
fits are tabulated in Table~\ref{tab:lifetimes}.  All the fits have acceptable 
quality.  The independent measurements are consistent with each other and
with the global lifetime fit.

We have made a detailed study of systematic effects using the charm data
itself, Monte-Carlo simulations, and a sample of $2\times10^6$ observed
$\kzii$ decays.  The non negligible contributions are tabulated in
Table~\ref{tab:systematics}.  The dominant contribution is the uncertainty in
the determination of the acceptance function.  This error was based on studies
of charm lifetime measurements for different targets, for different momentum
ranges, for different event multiplicities, for charm decays in different
z-regions, for varying sample-defining cuts, and for the use of proper time
instead of reduced proper time in the fit.

Many other effects, including mass reflections, effects of the presence of a
second charm particle in the event, interaction losses in the targets,
backgrounds induced by mismeasurements of charm decays, different fitting
techniques, different definitions of minimum distance $L_{min}$, etc., have
been studied.  Mass reflections were dominated by $D_s^+ - \Lambda_c^+$
reflection where the $K^+$ in $D_s^+ \rightarrow K^+K^-\pi^+$ decay was
misidentified as a proton.  The lifetime change with different choices of
sideband regions is negligible.  Decay tracks from the second charm particle
in the same event can influence the fit of the primary vertex and may lead to
an error in the distance $L$.  All small systematic errors were included
in the "other" entry of Table~\ref{tab:systematics}.

\begin{table}[!ht]
\centering
\begin{tabular}{|l||c|c|}
                        & $\Lambda_{c}^{+}$\hspace{1.0ex}$\Delta\tau$({\fsec}) 
                        & $D^{0}$\hspace{1.0ex}$\Delta\tau$({\fsec}) \\ \hline \hline
 acceptance             &           5.1    &           3.8    \\
 mass reflections       &           1.3    &                  \\
 background systematics &            -     &           1.0    \\
 second charm in event  &         $<1.0$   &         $<1.0$   \\
 other                  &         $<1.5$   &         $<1.5$   \\
\hline \hline                                                 
 total (quadrature)     &       \lcstau    &      \dzstau     \\
\end{tabular}
\caption{Systematic error contributions.}
\label{tab:systematics}
\end{table}

Based upon \lcnev $\Lambda_{c}^{+}$ and \dznev $D^{0}$ decays we observe
lifetimes of $\tau[\Lambda_{c}^{+}] = \lctau \pm \lcetau \pm \lcstau$~{\fsec}
and $\tau[D^{0}] = \dztau \pm \dzetau \pm \dzstau$~{\fsec}.  These results are
consistent with the present PDG averages ~\cite{PDG}: $\tau[\Lambda_{c}^{+}] =
206 \pm 12$~{\fsec} and $\tau[D^{0}] = 412.6 \pm 2.8$~{\fsec}.  The precision
of our $\tau[D^{0}]$ measurement is within a factor of 2 of the most precise
measurements~\cite{E687-D,E791-D,CLEO}.  As a final cross check we have
applied our analysis to $D^\pm\rightarrow K^\mp\pi^\pm\pi^\pm$ where our
acceptance corrections are much larger than in these analyses.  Our result
$\tau[D^\pm]=\dpkiitau\pm\dpkiietau$~{\fsec} (statistical error only) is
consistent with present PDG average~\cite{PDG} $\tau[D^\pm]=1051\pm
13$~{\fsec}.  The agreement with these precise measurements demonstrates our
control of systematic effects.  This $\tau[\Lambda_{c}^{+}]$ measurement has a
total error that is a factor of 2 smaller than the best previously-published
result~\cite{E687-L}. We look forward to measurements with similar precision
of the lifetimes of the other 3 stable charmed baryons, by us and others, in
the near future.  The set of precision lifetime measurements required for a
better understanding of charm weak decays should soon be available.

%
The authors are indebted to the staff of Fermi National Accelerator Laboratory
and for invaluable technical support from the staffs of collaborating
institutions.
This project was supported in part by Bundesministerium f\"ur Bildung, 
Wissenschaft, Forschung und Technologie, Consejo Nacional de 
Ciencia y Tecnolog\'{\i}a {\nobreak (CONACyT)},
Conselho Nacional de Desenvolvimento Cient\'{\i}fico e Tecnol\'ogico,
Fondo de Apoyo a la Investigaci\'on (UASLP),
Funda\c{c}\~ao de Amparo \`a Pesquisa do Estado de S\~ao Paulo (FAPESP),
the Israel Science Foundation founded by the Israel Academy of Sciences and 
Humanities, Istituto Nazionale di Fisica Nucleare (INFN),
the International Science Foundation (ISF),
the National Science Foundation (Phy \#9602178),
NATO (grant CR6.941058-1360/94),
the Russian Academy of Science,
the Russian Ministry of Science and Technology,
the Turkish Scientific and Technological Research Board (T\"{U}B\.ITAK),
the U.S. Department of Energy (DOE grant DE-FG02-91ER40664 and DOE contract
number DE-AC02-76CHO3000), and
the U.S.-Israel Binational Science Foundation (BSF).

\end{document}